\documentclass[,twocolumn,,trackchanges]{aastex631}
\usepackage{natbib}
\usepackage{amssymb}
\usepackage{amsmath}
\usepackage{epstopdf}

\usepackage{xcolor}
\usepackage[T1]{fontenc}
\usepackage{url}
\usepackage{epsfig}
\usepackage[mathscr]{euscript}
\usepackage{multirow}
\usepackage{threeparttable}

\begin{document}

\linespread{1.12}\selectfont
\title{Neutrino Emission from Gamma-ray Burst Jet inside the Cavity within Active Galactic Nucleus Accretion Disks}


\correspondingauthor{Kai Wang}
\email{kaiwang@hust.edu.cn}

\author[0009-0005-3856-7576]{Hao-Yu Yuan}
\affiliation{Department of Astronomy, School of Physics, Huazhong University of Science and Technology, Luoyu Road 1037, Wuhan, 430074, China}
\affiliation{College of Physics and Electronic Information Engineering, Guangxi Minzu Normal University, Chongzuo 532200, China}

\author[0009-0003-9792-9325]{Wen-Long Xu}
\affiliation{Department of Astronomy, School of Physics, Huazhong University of Science and Technology, Luoyu Road 1037, Wuhan, 430074, China}
\affiliation{Shiyan Campus, Hanjiang Normal University, No. 18, Beijing South Road, Shiyan 442000, Hubei, China}

\author[0000-0003-4976-4098]{Kai Wang}
\affiliation{Department of Astronomy, School of Physics, Huazhong University of Science and Technology, Luoyu Road 1037, Wuhan, 430074, China}
\affiliation{Hubei Key Laboratory of Gravitation and Quantum Physics, Huazhong University of Science and Technology, Wuhan, Hubei, China}
\affiliation{Hubei Fundamental Research Center for Physics, Wuhan, Hubei, China}

\author[0000-0003-3440-1526]{Wei-Hua Lei}
\affiliation{Department of Astronomy, School of Physics, Huazhong University of Science and Technology, Luoyu Road 1037, Wuhan, 430074, China}
\affiliation{Hubei Key Laboratory of Gravitation and Quantum Physics, Huazhong University of Science and Technology, Wuhan, Hubei, China}
\affiliation{Hubei Fundamental Research Center for Physics, Wuhan, Hubei, China}


\begin{abstract}
Active galactic nucleus (AGN) accretion disks are promising sites for compact binary mergers. Short gamma-ray burst (SGRB) jets from binary neutron star coalescence can propagate within low-density cavities carved by circumbinary outflows. We investigate the high-energy neutrino emission and detectability of such SGRB jets, adopting three seed photon components: prompt GRB emission, AGN disk thermal radiation, and external inverse Compton (EIC) scattered photons. Our results show that neutrino emission is dominated by proton-photon ($p\gamma$) interactions, with prompt GRB photons and AGN disk photons serving as the dominant target components. The inclusion of AGN disk photons significantly reshapes the neutrino spectrum, shifting the emission peak to lower energies. This effect becomes more pronounced as the mass of the central supermassive black hole decreases and the burst location approaches the black hole. For a fiducial model with a \(10^6\ M_\odot\) central black hole and the burst at 10 Schwarzschild radii, IceCube and IceCube-Gen2 achieve maximum luminosity distances of $\sim$180 Mpc and $\sim$400 Mpc, respectively. Next-generation high-sensitivity neutrino observatories, combined with multi-messenger observations, hold promise for identifying such GRBs in AGN environments.
\end{abstract}

\keywords{\href{https://astrothesaurus.org/uat/1100}{Neutrino astronomy (1100)}, \href{https://astrothesaurus.org/uat/629}{Gamma-ray bursts (629)}, \href{http://astrothesaurus.org/uat/562}{Galaxy accretion disks (562)}}


\section{Introduction}\label{Intro}

The detection of the binary neutron star merger GW170817 \citep{Abbott-et-al.(2017)}, together with the gamma-ray burst GRB 170817A \citep{Abbott-etal.(2017)848, Goldstein-et-al.(2017), Savchenko-et-al.(2017), Zhang-et-al.(2018), Alexander-et-al.(2017)848, Hallinan-et-al.(2017)358, Margutti-et-al.(2017)848, Lazzati-et-al.(2018)120} and the kilonova AT2017gfo \citep{Abbott-et-al.(2017)848L13, Arcavi-et-al.(2017)551, Andreoni-et-al.(2017)34, Chornock-et-al.(2017)848, Covino-et-al.(2017)1, Hu-et-al.(2017)62, Yu-et-al.(2018)861}, inaugurated the era of multi-messenger astronomy. Such binary compact object (BCO) mergers are generally expected to occur in low-density environments (e.g., interstellar medium). The formation channels for BCOs are still subject to debated. Some formation channels, however, suggest that BCO mergers may also take place in high-density environments \citep{Zhu-Chen(2025), Bartos-et-al.(2017)}, such as active galactic nucleus (AGN) accretion disks.

AGNs are widely believed to be powered by the accretion of dense material from the surrounding disk onto the central supermassive black hole \citep[SMBH,][]{Lynden-Bell(1969), Shakura-Sunyaev(1973)}. The dense gas in AGN disks can promote the formation and merger of BCOs \citep{Zhu-Chen(2025), Bartos-et-al.(2017)}. Recently, the LIGO and Virgo collaborations reported the detection of the binary black hole (BBH) merger event GW190521 \citep{Abbott-et-al.(2020)}. Although BBH mergers are not expected to produce electromagnetic (EM) counterparts in typical low-density environments \citep{McKernan-et-al.(2019)884}, the Zwicky Transient Facility (ZTF) reported a candidate optical transient in the AGN J14942.3+344929, ZTF19abanrhr, potentially associated with GW190521 \citep{Graham-et-al.(2020)} $\sim$34 days after the merger. Subsequently, \citet{Graham-et-al.(2023)} analyzed BBH merger events detected by LIGO–Virgo during O3 and found nine potentially associated EM sources, based on the ZTF sky coverage, AGN information, and merger geometry. More recently, \citet{He-et-al.(2025)} found six plausible optical flares potentially associated with GW231123 using a similar method. Although not yet confirmed, these GW-EM associations suggest that AGN disks might provide an important channel for the formation of BCOs \citep{McKernan-et-al.(2020)498}.

Stars in AGN disks may develop through two channels: gravitationally unstable collapse in outer regions \citep{Paczynski(1978)28,Goodman(2003),Dittmann-Miller(2020)493}, or capture from surrounding stellar clusters \citep{Artymowicz-et-al.(1993)409,Fabj-et-al.(2020)499}.
The death of massive stars leaves behind compact objects, such as neutron stars (NSs) and black holes (BHs).
Compact objects undergo migration driven by gas torques in AGN disks, and accumulate near migration traps, where the total torque acting on them becomes zero, promoting the formation of BCOs \citep{Secunda-et-al.(2019),McKernan-et-al.(2020)498,Tomar-et-al.(2026),Rowan-et-al.(2025)}.
GRB jets can be produced either by massive stellar collapse or by BCO (e.g., NS-NS and NS-BH) mergers, corresponding to long GRBs (LGRBs) or short GRBs (SGRBs) in AGN disks, respectively. 
\citet{ Lazzati-e-tal.(2022)} showed that interaction with the AGN disk gas can significantly prolong the duration of SGRBs. In addition,
\citet{Zhu-et-al.(2021d)} found that SGRB jets are likely to be choked by the AGN disk medium, and only thermal X-ray emission powered by cocoon shock breakout can be observed \citep{Zhang-et-al.(2024)}. \citet{Perna-et-al.(2021)} argued that jet breakout becomes harder with increasing SMBH mass. For LGRBs, lower progenitor metallicity tends to produce longer-lived jets, enhancing their ability to break out of the AGN disk \citep{Wei-et-al.(2025)}.
Recently, GRB 191019A has been suggested as a possible GRB originating from an AGN disk environment \citep{Levan-et-al.(2023),Lazzati-et-al.(2023)}, providing further evidence that AGN disks may serve as sites of BCO mergers and multi-messenger transients.

High-energy neutrinos, which benefit from the negligible absorption, can serve as unique messengers to probe the activities occurring within the AGN disk.
GRB jets within the AGN disk can be natural ``hidden'' cosmic-ray accelerators
\citep{PhysRevLett.116.071101,2025arXiv251106707M}. 
\citet{Zhu-et-al.(2021b)} found that the choked GRB jet in the AGN disk can produce TeV-PeV neutrinos through the reaction between protons and jet head photons, and the neutrino emission may be detected by IceCube.
The interaction between the jet from BBH mergers and the AGN disk can create several shocks, i.e., the forward shock, reverse shock, internal shock, and collimation shock.
\citet{Zhou-Wang(2023)} found that the forward shock cannot efficiently accelerate protons. The higher energy neutrinos ($\gtrsim$ PeV) are contributed by reverse and collimation shocks, but the lower energy neutrinos are only due to the reverse shock.
\citet{Zhu(2024)} found that IceCube can detect the neutrinos from GW190521 if the BBH merger arises in the outer disk region.
\citet{Tagawa-et-al.(2023)} argued that the neutrino emission of NGC 1068 may come from internal shocks of the jets in AGN disks.
The evolution of BCOs can drive strong winds \citep{Kimura-et-al.(2021)}. On the one hand, the interaction between the wind and the AGN disk can produce neutrino emission \citep{Ma-Wang(2024),Zhou-et-al.(2023),Zhu-et-al.(2021a)}; on the other hand, the wind can blow away the AGN disk gas around the BCO, forming a low-density cavity.
If the cavity has not been refilled with the surrounding gas and the jet is oriented away from the mid-plane of the AGN disk, the GRB jet can successfully break out \citep{Kimura-et-al.(2021),Yuan-et-al.(2022), Yuan-et-al.(2025)}.

Although the successful jet can be observed when the cavity is formed, the EM emission of the jet would be affected by AGN disk photon fields, especially GeV emission of the jet \citep{Yuan-et-al.(2022)}. In addition, the AGN disk photon fields can also mediate proton acceleration and cooling. \citet{Yuan-et-al.(2022)} qualitatively discussed how AGN disk photon fields can influence the efficiency of neutrino emission, but they did not perform a detailed analysis of the radiation properties of neutrinos. A detailed investigation of neutrino emission across different energy bands can provide an important diagnostic for probing the physical conditions of GRBs in AGN disks and may offer a means of identifying such events through their neutrino signatures.

In this work, we include the cavity created before jet launch and focus on the impact of the AGN disk photon field on neutrino radiation within the internal dissipation region of the SGRB jet.
We assume BCO mergers are located at the mid-plane of the AGN disk.
Moreover, the jet, which is perpendicular to the mid-plane, has a uniform structure, and the line of sight is on-axis.
The paper is organized as follows. In Section \ref{Sec: model}, we introduce the model framework and ingredients; 
the results of the numerical calculation are displayed in Section \ref{Sec:results}. Conclusions and discussions are presented in Section \ref{sec: conclusion}.
Throughout the paper, primed quantities denote those in the jet comoving frame, and unprimed quantities refer to the observer frame.

\section{model overview}\label{Sec: model}
Outflows driven by accretion feedback from BCOs embedded in the AGN disk can excavate cavities within the disk. Once accretion ceases, these cavities are refilled by the surrounding disk material. When the jet is launched, if the cavity size is comparable to the scale height of the AGN disk and the jet is directed nearly perpendicular to the disk midplane, the jet propagation will be unaffected by the AGN disk material.
When the jet traverses the cavity, the electron energy spectra and electromagnetic radiation characteristics in the internal dissipation region will be altered by the external inverse Compton (EIC) process involving AGN disk photon fields and the jet electrons \citep{Yuan-et-al.(2022)}.
We consider three types of seed photons when analyzing the neutrino emission from the dissipation region within the jet, namely, prompt photons in the internal dissipation region, AGN disk photons, and EIC photons.
A schematic diagram of our model is shown in Fig.~\ref{fig:model_schematic}.

In this section, we first present the number density of seed photons for the p$\gamma$ interaction. We then discuss the cooling processes of protons and the production of neutrinos.

\subsection{Seed photons for p$\gamma$ Interactions}
\subsubsection{AGN disk model and its thermal photons}
\begin{figure*}
\centering
\includegraphics [angle=0,scale=0.25] {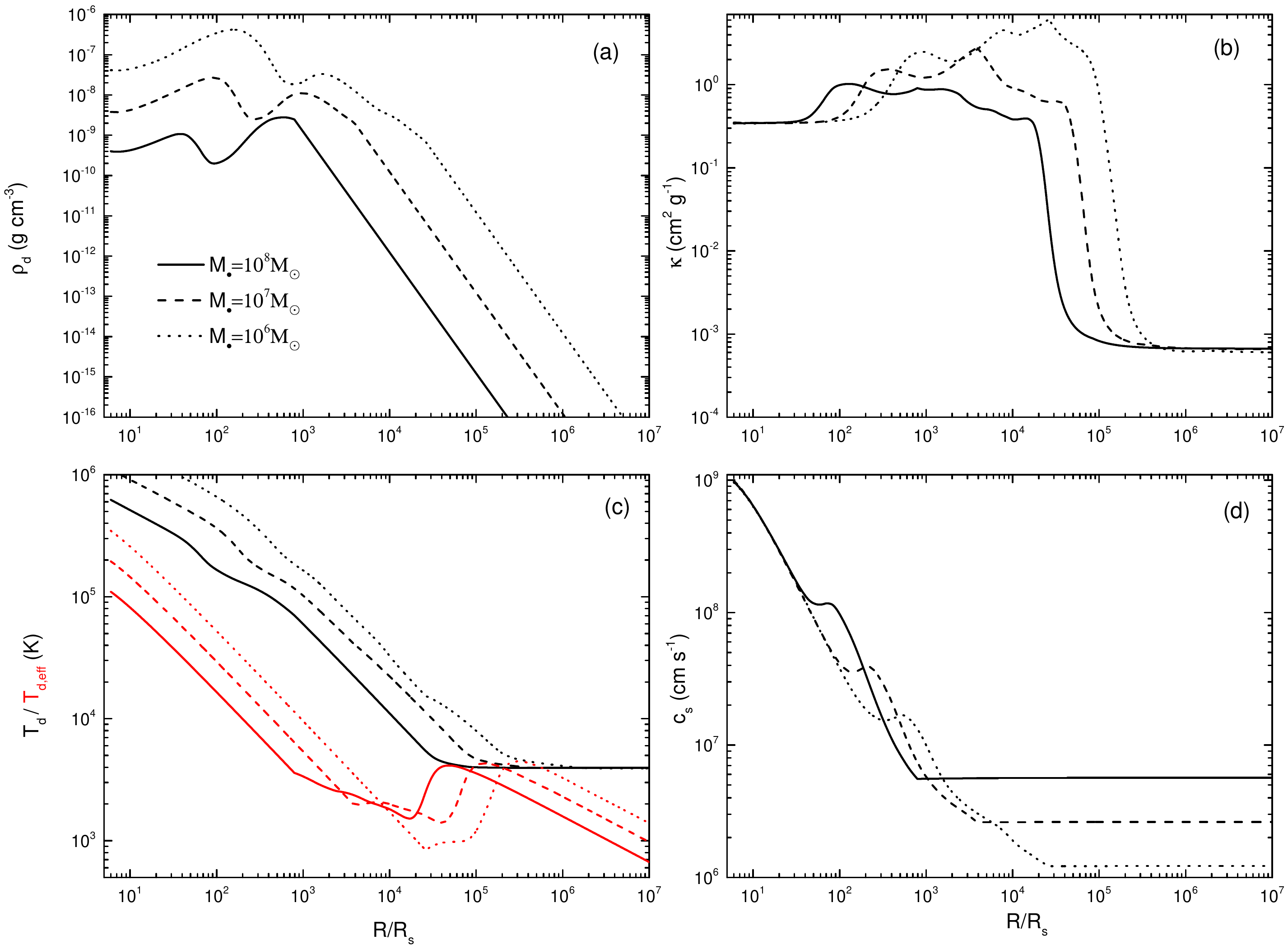}
\caption{AGN disk profiles for the density $\rho_{\rm d}$, opacity $\kappa_{\rm d}$, mid-plane temperature $T_{\rm d}$, effective temperature $T_{\rm eff}$ and sound velocity $c_{\rm s}$ with $M_\bullet=10^6M_\odot$ (dotted line), $10^7M_\odot$ (dashed line) and $10^8 M_\odot$ (solid line).
$R_{\rm s}=2GM_\bullet/c^2$ is the Schwarzschild radius.
}
\label{fig:AGNdiskprofile_1}
\end{figure*}

\begin{figure*}
\centering
\includegraphics [angle=0,scale=0.25] {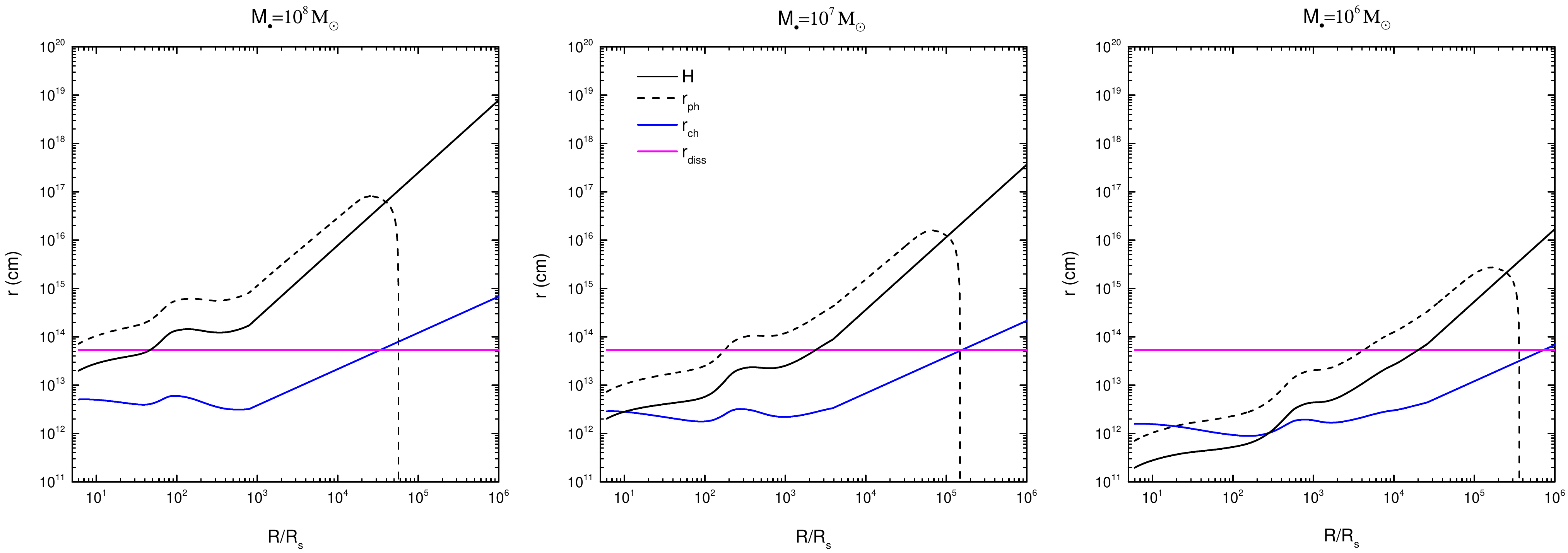}
\caption{The disk scale height ($H$, solid black line), the disk photosphere radius ($r_{\rm ph}$, dashed black line), the jet choked radius ($r_{\rm ch}$, blue line), and the internal dissipation radius ($r_{\rm diss}$, magenta line) evolve with $R$.
Note that $r_{\rm ch}$ is defined after the cavity has been refilled.
}
\label{fig:AGNdiskprofile_2}
\end{figure*}

In this paper, we adopt the AGN disk model developed by \citet{Sirko-Goodman(2003)}, which is a modification of the $\alpha$ prescription disk model \citep{Shakura-Sunyaev(1973)}, with a constant accretion rate.
We use the Python module \textit{pAGN}, developed by \citet{Gangardt-et-al.(2024)}, to calculate the radial profile of the AGN disk.
In the following calculations, we use $\alpha=0.01$ and the total luminosity of the disk $L_0=\epsilon_{\rm d}\dot{M}_\bullet c^2=0.5L_{\rm Edd}$, where $\dot{M}_\bullet$ is the accretion rate of the SMBH, $\epsilon_{\rm d}=0.1$ is the conversion efficiency from accretion to radiation, and $c$ is the speed of light. $L_{\rm Edd}=4\pi GM_\bullet m_{\rm p}c/\sigma_{\rm T}$ is the Eddington luminosity, where $m_{\rm p}$ is the proton mass and $\sigma_{\rm T}$ is the
Thomson cross section.
The radial profiles of the density $\rho_{\rm d}$, opacity $\kappa_{\rm d}$, mid-plane temperature $T_{\rm d}$ (or effective temperature $T_{\rm eff}$), and sound velocity $c_{\rm s}$ are shown in Fig. \ref{fig:AGNdiskprofile_1} for the central SMBH mass $M_\bullet=10^6M_\odot$, $10^7M_\odot$, and $10^8M_\odot$.
The scale height of the disk $H$ and the photosphere $r_{\rm ph}$ are shown in Fig. \ref{fig:AGNdiskprofile_2} with solid and dashed black lines.
The photospheric radius varies across the AGN disk. In the inner disk, it is slightly above the disk scale height, whereas in the outer disk it shrinks rapidly because of a steep decrease in both the opacity and the gas density.

\begin{figure*}
\centering
\includegraphics [angle=0,scale=0.25] {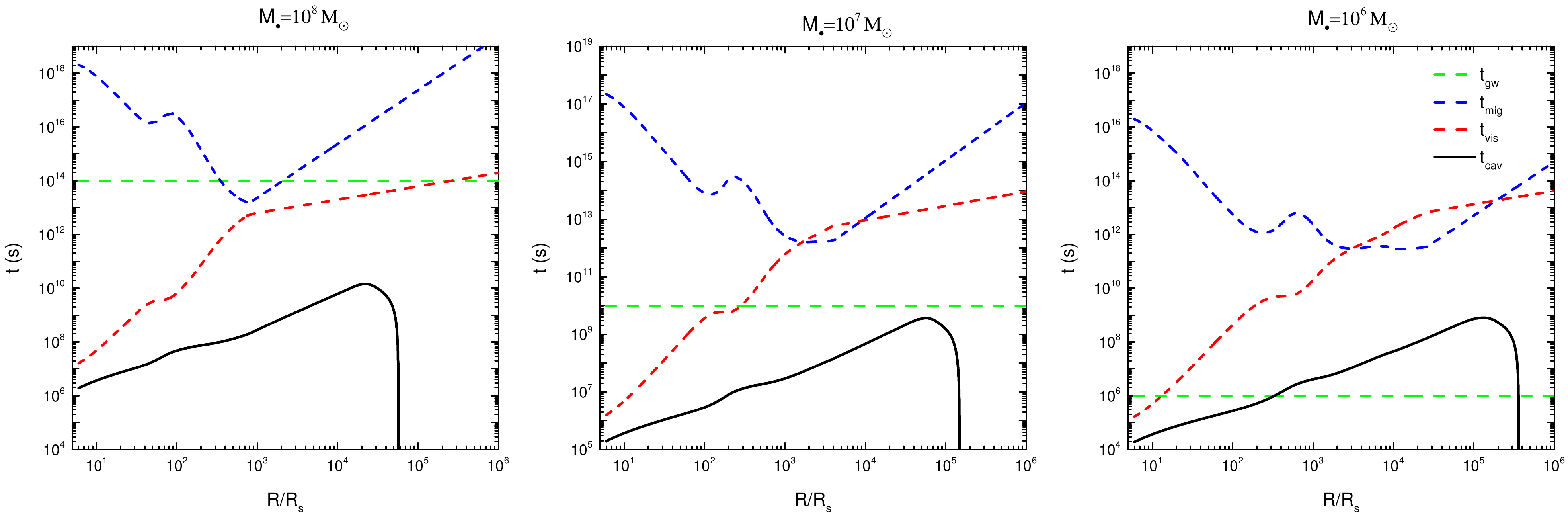}
\caption{The timescale of BCO merger $t_{\rm gw}$ (green line), migration (blue line), AGN disk viscosity (red line), and cavity formation $t_{\rm cav}$ for $r_{\rm w}=r_{\rm ph}$ (black line) evolve with $R$.
}
\label{fig:cavity_timescale}
\end{figure*}

\begin{figure}
\centering
\includegraphics [angle=0,scale=0.45] {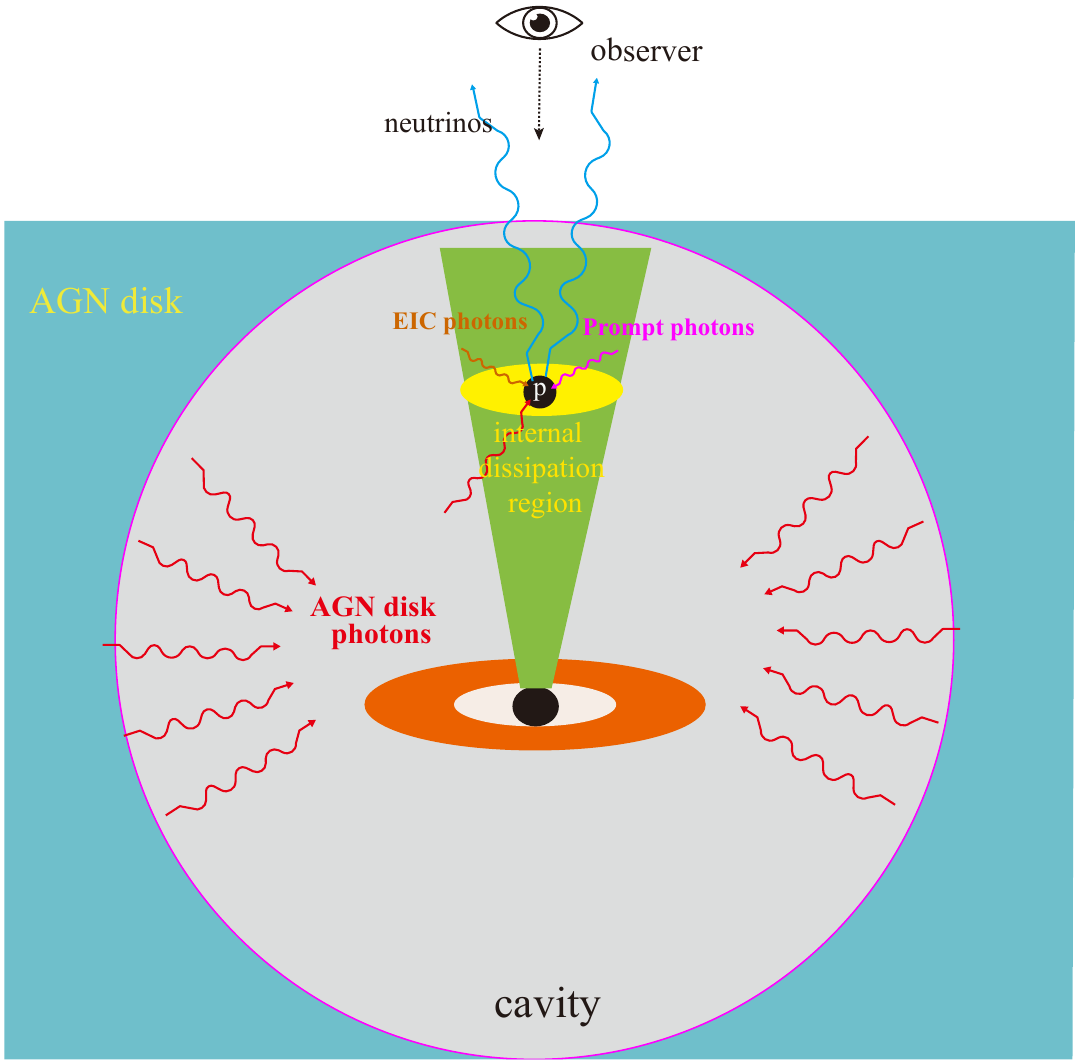}
\caption{A schematic diagram of the jet launched by the BCO merger embedded in the AGN disk. The cavity is created by the powerful outflow during the BCO evolution.
The regions of the AGN disk and the cavity are represented by the light blue and light gray shaded regions, respectively. The GRB jet is shown by the light green cone. 
There are three types of seed photons: AGN disk photons (red rays), AGN disk photons after undergoing external inverse Compton (EIC) with the electrons inside the jet (orange rays, referred to as EIC photons), and prompt photons (magenta rays).
The blue rays are the neutrino emission from the p$\gamma$ reactions. 
}
\label{fig:model_schematic}
\end{figure}

BCOs embedded in AGN disks undergo super-Eddington accretion, with the accretion rate $\dot{M}_{\rm BCO}=\eta_{\rm BCO}\dot{M}_\bullet$, where $\eta_{\rm BCO}\sim 0.1$ is the ratio of the BCO accretion rate to the SMBH accretion rate \citep{Kimura-et-al.(2021)}.
Accretion feedback can drive a powerful outflow \citep{Ohsuga-et-al.(2009), Jiang-et-al.(2014), Sadowski-et-al.(2014),Tagawa-et-al.(2023)950,Tagawa-et-al.(2022)}. We assume that the outflow is spherical and that the equation of motion for bubble expansion in the AGN disk is $r_{\rm w}\approx 0.88(L_{\rm w}t^3/\rho_{\rm BCO})^{1/5}$ \citep{Weaver-et-al.(1977), Koo-McKee(1992)}, where $r_{\rm w}$ is the bubble radius, $L_{\rm w}=\eta_{\rm w}\dot{M}_{\rm BCO}v_{\rm w}^2$ is the outflow luminosity, $v_{\rm w}\sim 10^9~ \rm{cm}~\rm s^{-1}$ is the outflow velocity, and $\rho_{\rm BCO}$ is the mid-plane density at the position of the BCO. $\eta_{\rm w}$ is the radiation efficiency, and we adopt $\eta_{\rm w}\sim 0.3$ as a fiducial value. The density of the outflow is $\rho_{\rm w}=\dot{M}_{\rm BCO}/(4\pi r^2 v_{\rm w})$, where $r$ is the distance from the BCO at the direction of perpendicular to the mid-plane. The photosphere radius of the outflow $r_{\rm ph,w}$ can be defined by $\int_{r_{\rm ph,w}}^\infty \kappa_{\rm w}\rho_{\rm w}dr=1$, where $\kappa_{\rm w}\approx 0.4$ is the opacity of the outflow. One can derive that $r_{\rm ph,w}\approx 2\times 10^{11}~\frac{M_\bullet}{10^6M_\odot}~ {\rm cm}$.
Equating $r_{\rm w}$ to $r_{\rm ph}$, one can derive the timescale for the cavity to reach the surface of the AGN disk, $t_{\rm cav}$,
\begin{equation}
\label{Eq:cavity_timescale}
t_{\rm cav}=1.24\left(\frac{\rho_{\rm BCO}r_{\rm ph}^5}{L_{\rm w}}\right)^{1/3}.
\end{equation}
For a cavity to form before the merger of the compact binary, the cavity formation timescale should satisfy $t_{\rm cav}\lesssim \min (t_{\rm gw}, t_{\rm vis}, t_{\rm mig})$ \citep{Kimura-et-al.(2021)}, and these timescales are defined as follows
\begin{eqnarray}
\label{Eq:timescale}
\nonumber t_{\rm gw}=\frac{5}{128}\frac{\dot{m}_{\rm BCO}^4}{A_{\rm in}^4}\frac{GM_{\rm BCO}}{c^3},\\
t_{\rm vis}=\frac{R^3}{\alpha H^2 v_{\rm K}},\\
\nonumber t_{\rm mig}=\frac{HM_\bullet^2}{2M_{\rm BCO}Rv_{\rm K}\rho_{\rm d}},
\end{eqnarray}
where $t_{\rm gw}$, $t_{\rm vis}$, and $t_{\rm mig}$ are timescales of BCO merger, AGN
disk viscosity, and migration, respectively. 
$M_{\rm BCO}=5M_\odot$ is the mass of the BCO, $\dot{m}_{\rm BCO}$ is the ratio of BCO accretion to the Eddington accretion rate, $R$ is the distance between the SMBH and the BCO, and $v_{\rm K}=\sqrt{GM_\bullet/R}$ is the Kepler velocity.
$A_{\rm in}\sim 10$ is the ratio between the inner edge of the
BCO disk and its orbital major axis \citep{Nixon-et-al.(2013)}.
The evolution of these timescales with $R$ is shown in Fig. \ref{fig:cavity_timescale},
which shows that a cavity can be fully formed for $M_\bullet =10^7M_\odot$ and $10^8 M_\odot$. However, for less massive SMBHs (e.g., $M_\bullet=10^6M_\odot$), the cavity does not fully develop for $R\gtrsim 300 R_{\rm s}$.

BCOs are typically located in the inner regions of AGN disks \citep{Bellovary-et-al.(2016), Tagawa-et-al.(2020)}, where the optical depth is much larger than unity. Therefore, the local photon number density for the cavity can be simply approximated with a blackbody spectrum \citep[in units of ${\rm erg^{-1}~cm^{-3}}$,][]{Yuan-et-al.(2022)},
\begin{equation}
\label{Eq:AGN_photon_density}
n_{\rm d}(\varepsilon_\gamma)=\frac{8\pi}{(hc)^3}\frac{\varepsilon_\gamma^2}{\exp{(\varepsilon_\gamma/k_{\rm B}T)}},
\end{equation}
where $\varepsilon_\gamma$ is the photon energy, $k_{\rm B}$ is the Boltzmann constant, $h$ is the Planck constant, and $T$ is the local temperature. For simplicity, unless otherwise specified, we adopt $T=T_{\rm d,eff}$ throughout the following calculations\footnote{Following \citet{Yuan-et-al.(2022)}, we assume that the effective disk temperature provides the dominant source of seed photons for the p$\gamma$ interaction. Nevertheless, thermal photons originating from the hotter, deeper layers of the AGN disk, as well as those produced by the circumbinary disk, may modify the seed photon field. We will discuss the effects of these components on the neutrino emission in the final section.}.

\subsubsection{Prompt emission of internal dissipation}
In this work, we focus on the jet dissipation process in the dissipation radius ($r_{\rm diss}$), including internal shocks \citep{Rees-Meszaros(1994)} and magnetic reconnection \citep{McKinney-Uzdensky(2012)}. The emission of internal dissipation can be described by the Band component \citep{Band-et-al.(1993)}. 
The presence of the synchrotron radiation death line implies that synchrotron emission cannot fully explain the Band component in GRB prompt emission \citep{Preece-et-al.(1998),Meszaros-Rees(2000),Zhang(2018)}.
Therefore, we adopt the empirical Band function to describe the photon number spectrum of the jet emission in the internal dissipation region, which can be expressed as \citep{Band-et-al.(1993)},
\begin{equation}
\label{eq: band}
n'_{\rm band}(\varepsilon_{\gamma}')\propto\left\{
\begin{aligned}
&\varepsilon_{\gamma}'^\alpha\exp \left(-\frac{\varepsilon_{\gamma}'}{\varepsilon_{\gamma,0}'}\right), &\varepsilon_{\gamma}'<(\alpha-\beta)\varepsilon_{\gamma,0}' \\
\\
&\left[(\alpha-\beta)\varepsilon_{\gamma,0}'\right]^{\alpha-\beta}\times\\
&\exp (\beta-\alpha)\varepsilon_{\gamma}'^\beta, &\varepsilon_{\gamma}'>(\alpha-\beta)\varepsilon_{\gamma,0}'
\end{aligned}
\right. ,
\end{equation}
where $\varepsilon'_{\gamma,0}=\varepsilon'_{\gamma,p}/(2+\alpha)$ is the break energy and $\varepsilon'_{\rm \gamma,p}$ is the peak energy of the spectrum. $\alpha$ and $\beta$ are the low- and high-energy photon spectral indices, e.g., $\alpha=-1$ and $\beta=-2$ \citep{Preece-et-al.(2000)}. The normalization factor in Eq.\ref{eq: band} can be obtained by $\epsilon_{\rm e}L_{\rm j,iso}/4\pi c \Gamma_{\rm j}^2r_{\rm diss}^2=\int_{\varepsilon_{\rm \gamma,m}'}^{\varepsilon_{\rm \gamma,M}'}d\varepsilon_\gamma' \varepsilon_\gamma'n'_{\rm band}(\varepsilon_{\gamma}')$, where $\epsilon_{\rm e}$ is the fraction of jet kinetic energy that is converted to electrons, $L_{\rm j,iso}$ is the isotropic kinetic luminosity of the jet, $\Gamma_{\rm j}$ is the bulk Lorentz factor of the jet, $\varepsilon_{\rm \gamma,m}'$ and $\varepsilon_{\rm \gamma,M}'$ are the minimum and maximum photon energies.
In addition, we derive the  Comptonized disk photon spectrum in Sec. \ref{Sec: electron}.
By using the photon spectrum in the dissipation region and a given proton spectrum, we can obtain the neutrino radiation fluence (see Sec. \ref{Sec: neutrino}).

Notably, if the jet is choked before reaching the dissipation radius, there would be no prompt emission to provide seed photons while the cavity is being refilled.
Adopting the methodology from \citet{Zhu-et-al.(2021d)}, we determine the choked radius $r_{\rm ch}$, defined as the radius at which the jet tail catches up with the jet head, and reveal that it exceeds the internal dissipation radius only for BCOs in the outer AGN disk, as shown in Fig. \ref{fig:AGNdiskprofile_2}. Because BCOs predominantly originate in the inner disk, a cavity is necessary in our model. 
Furthermore, once accretion halts, the AGN disk medium replenishes the cavity \citep{Wang-et-al.(2021a), Chen-et-al.(2023),Tagawa-et-al.(2022),Tagawa-et-al.(2023)946}, potentially leading to jet choking before internal dissipation takes place. We henceforth operate under the assumption that the cavity is established when the jet is launched.

\subsubsection{EIC photons}\label{Sec: electron}
Electrons can be continuously accelerated at internal dissipation shocks. To obtain the electron spectral distribution of the radiation region, we solve the steady-state electron continuity equation,
\begin{equation}
\label{Eq:N_gamma_e}
\frac{N_{\rm \gamma_{\rm e}'}'}{t_{\rm dyn}'}-\frac{\partial}{\partial \gamma_{\rm e}'}\left(\dot{\gamma}_{\rm e}'N_{\rm \gamma_{\rm e}'}'\right)=\dot{Q}_{\rm e,inj}',
\end{equation}
where $\gamma_{\rm e}'$ is the electron Lorentz factor, $N_{\rm \gamma_{\rm e}'}'=dN_{\rm e}'/d\gamma_{\rm e}'$ is the spectrum of electron energy, $t_{\rm dyn}'=r_{\rm diss}/\Gamma_{\rm j}c$ is the dynamical timescale of the jet, which can represent adiabatic losses or the escape timescale, and $\Gamma_{\rm j}$ is the bulk Lorentz factor of the jet.
$\dot{\gamma}_{\rm e}'=\gamma_{\rm e}'/t_{\rm e,cool}'$ is the energy loss rate of electrons and $t_{\rm e,cool}'$ is the cooling timescale of electrons.
$\dot{Q}_{\rm e,inj}'$ is the energy injection rate of electrons, and we consider a power-law distribution with a slope $s$, e.g., $\dot{Q}_{\rm e,inj}'\propto (\gamma'_{\rm e}/\gamma'_{\rm e,m})^{-s}$ \citep{Zhang-et-al.(2019)}.
Assuming the index $s=2.2$, we can obtain the normalization factor for the injection term can be obtained by $\int_{\gamma_{\rm e,m}'}^\infty d\gamma_{\rm e}'\gamma_{\rm e}'m_{\rm e}c^2\dot{Q}_{\rm e,inj}'=\epsilon_{\rm e}L_{\rm j,iso}/\Gamma_{\rm j}^2$, where $m_{\rm e}$ is the mass of the electron, and $\gamma_{\rm e,m}'=\frac{s-2}{s-1}\frac{\epsilon_{\rm e}(\Gamma_{\rm j}-1)m_{\rm p}}{m_{\rm e}}$ is the minimum Lorentz factor of electrons.
Within the internal dissipation region, electrons are accelerated by shocks with an acceleration timescale of $t_{\rm e,acc}'=\gamma_{\rm e}'m_{\rm e}c/eB'$, where $e$ is the electron charge, $B'=\sqrt{2\epsilon_{\rm B}L_{\rm j,iso}/r_{\rm diss}^2\Gamma_{\rm j}^2c}$ is the strength of the magnetic field and $\epsilon_{\rm B}$ is the fraction of jet kinetic energy that is converted to magnetic field energy.

Accelerated electrons lose energy via different mechanisms \citep{Bosnjak-et-al.(2009)}, e.g., synchrotron emission, inverse Compton (IC) process, and adiabatic cooling. The cooling rates of the electron for these mechanisms can be denoted by $\dot{\gamma}_{\rm e,syn}'$, $\dot{\gamma}_{\rm e,IC}'$ and $\dot{\gamma}_{\rm e, adi}'$, which can be expressed as \citep{Jones(1968), Blumenthal-Gould(1970), Fan-et-al.(2008), Uhm-et-al.(2012), Geng-et-al.(2018)},
\begin{equation}
\label{Eq:dot_gamma_e_syn}
\dot{\gamma}_{\rm e, syn}'=\frac{\sigma_{\rm T}B'^2\gamma_{\rm e}'^2}{6\pi m_{\rm e}c},
\end{equation}
\begin{eqnarray}
\label{Eq:dot_gamma_e_IC}
\nonumber\dot{\gamma}_{\rm e, IC}'=&\frac{1}{m_{\rm e}c^2}\frac{3\sigma_{\rm T}c}{4\gamma_{\rm e}'^2}\int_{\nu_{\rm min}'}^{\nu_{\rm max}'}\frac{n'(\nu')d\nu'}{\nu'}\\
&\times\int_{\nu_{\rm IC,min}'}^{\nu_{\rm IC,max}'}h\nu_{\rm IC}'d\nu_{\rm IC}'F(q,g),
\end{eqnarray}
\begin{equation}
\label{Eq:dot_gamma_e_adi}
\dot{\gamma}_{\rm e, adi}'=\gamma_{\rm e}'/t_{\rm dyn}'
\end{equation}
the corresponding cooling timescales are presented by $t_{\rm e,syn}'$, $t_{\rm e, IC}'$ and $t_{\rm e,adi}'$, respectively.
$\sigma_{\rm T}$ is the Thomson cross section. $\nu'$ is the frequency, and $n'(\nu')$ is the number density of the seed photons for a specific frequency, and there are two different sources of the seed photons: electron synchrotron photon fields ($n_{\rm syn}'$) and AGN disk photon fields ($n_{\rm d}'$).
In the calculation, we consider two IC processes, namely synchrotron self-Compton (SSC) and EIC scattering. Their cooling rates (timescales) are denoted by $\dot{\gamma}_{\rm e,SSC}'$ and $\dot{\gamma}_{\rm e,EIC}'$ ($t_{\rm e,SSC}'$ and $t_{\rm e,EIC}'$), respectively.
$F(q,g)=2q\ln q+(1+2q)(1-q)+\frac{(4qg)^2}{2(1+4qg)}(1-q)$, where $g=\frac{\gamma_{\rm e}'h\nu'}{m_{\rm e}c^2}$, $w=\frac{h\nu_{\rm IC}'}{\gamma_{\rm e}'m_{\rm e}c^2}$ and $q=\frac{w}{4g(1-w)}$. The lower and upper limits of $\nu_{\rm IC}'$ are $\nu_{\rm IC,min}'=\nu'$ and $\nu_{\rm IC,max}'=\frac{\gamma_{\rm e}'m_{\rm e}c^2}{h}\frac{4g}{4g+1}$, respectively. For simplicity, the secondary IC is not considered here.
Therefore, the total cooling rate of the electron is
\begin{equation}
\label{Eq:t_e_cool}
\dot{\gamma}_{\rm e,cool}'=\dot{\gamma}_{\rm e,adi}'+\dot{\gamma}_{\rm e,syn}'+\dot{\gamma}_{\rm e,SSC}'+\dot{\gamma}_{\rm e,EIC}',
\end{equation}

The synchrotron radiation power for a specific $\nu'$,
\begin{equation}
\label{Eq:P_syn}
P_{\rm syn}'(\nu')=\frac{\sqrt{3}e^3B'}{m_{\rm e}c^2}\int_{\gamma_{\rm e,min}'}^{\gamma_{\rm e,max}'}N_{\gamma_{\rm e}'}'F(\nu'/\nu_{\rm c}')d\gamma_{\rm e}',
\end{equation}
is given by the above expression and the number density of synchrotron photons is
\begin{equation}
\label{Eq:n_syn}
n_{\rm syn}'(\nu')=\frac{1}{4\pi r_{\rm diss}^2}\frac{P_{\rm syn}'(\nu')}{ch\nu'},
\end{equation}
where $\nu_{\rm c}'=3eB'\gamma_{\rm e}'^2/(4\pi m_{\rm e}c)$ is the characteristic frequency, $F(\nu'/\nu_{\rm c}')=\nu'/\nu_{\rm c}'\int_{\nu'/\nu_{\rm c}'}^\infty K_{5/3}(\xi)d\xi$ and $K_{5/3}(\xi)$ is the modified Bessel function of order $5/3$.
The power of the IC process can be expressed as \citep{Blumenthal-Gould(1970), Fan-et-al.(2008)},
\begin{eqnarray}
\label{Eq:P_IC}
\nonumber
P_{\rm IC}'(\nu'_{\rm IC})=&\frac{3\sigma_{\rm T}ch\nu_{\rm IC}'}{4}\int_{\nu_{\rm min}'}^{\nu_{\rm max}'}\frac{n'(\nu')d\nu'}{\nu'}\\
&\times \int_{\gamma_{\rm e,min}'}^{\gamma_{\rm e,max}'}\frac{F(q,g)}{\gamma_{\rm e}'^2}N_{\gamma_{\rm e}'}'d\gamma_{\rm e}',
\end{eqnarray}
and the photon number density is,
\begin{equation}
\label{Eq:n_IC}
n_{\rm IC}'(\nu'_{\rm IC})=\frac{1}{4\pi r_{\rm diss}^2}\frac{P_{\rm IC}'(\nu'_{\rm IC})}{ch\nu'_{\rm IC}}.
\end{equation}
By substituting $n_{\rm syn}'$ and $n_{\rm d}'$ into Eq.\ref{Eq:n_IC}, we obtain the photon number densities for SSC and EIC processes, respectively.
The jet propagating in the cavity undergoes high-energy photon attenuation due to pair production with low-energy photons from the AGN disk's radiation field (i.e., $\gamma\gamma$ attenuation) when $\varepsilon'\varepsilon_{\rm seed}'\gtrsim (m_{\rm e}c^2)^2$, where $\varepsilon'$ is the high-energy photon energy and $\varepsilon_{\rm seed}'$ is the low-energy target photon energy. The optical depth of $\gamma\gamma$ process is \citep{ZhangBT-et-al.(2021), Murase-et-al.(2011)},
\begin{eqnarray}
\label{Eq:tau_gg}
\tau_{\rm \gamma\gamma}=H/2\int_{-1}^{1}d\mu(1-\mu)\int d\varepsilon'n_{\rm d}'(\varepsilon')\sigma_{\rm \gamma\gamma}(S),
\end{eqnarray}
and the annihilation cross section is,
\begin{eqnarray}
\nonumber
\sigma_{\rm \gamma\gamma}(S)=&\frac{3}{16}\sigma_{\rm T}\left(1-\beta_{\rm cm}^2\right)\times\\
\nonumber
&\left[\left(3-\beta_{\rm cm}^4\right)\ln \frac{1+\beta_{\rm cm}}{1-\beta_{\rm cm}}-2\beta_{\rm cm}\left(2-\beta_{\rm cm}^2\right)\right],
\end{eqnarray}
where $\beta_{\rm cm}=\sqrt{1-4/S}$ and $S=2\varepsilon'\varepsilon_{\rm seed}'(1-\mu)$ is the Mandelstam variable.

\subsection{proton acceleration and cooling}\label{Sec: neutrino}
We use $n'_\gamma(\varepsilon'_\gamma)$ to express the photon density in a dissipation region for a specific photon energy $\varepsilon'_\gamma$, which is the seed photon energy for the p$\gamma$ process. We adopt the minimum and maximum photon energy as $\varepsilon'_{\gamma,m}=0.1$ eV and $\varepsilon'_{\gamma,M}=10^6$ eV, respectively, as in \citet{Murase-Nagataki(2006a)}.
Besides, the specific proton energy distribution can be expressed as 
\begin{equation}
\label{Eq:n_p}
n_{\rm p}(\varepsilon_p)\propto \varepsilon_p^{-2},
\end{equation}
where $\varepsilon_p$ is the proton energy. The total energy of the protons is $E_{\rm p,iso}=\epsilon_{\rm p}E_{\rm j, iso}$, where $\epsilon_{\rm p}$ is the fraction of jet kinetic energy that is converted to protons, $E_{\rm j, iso}=L_{\rm j,iso}t_{\rm j}$ is the isotropic kinetic energy and $t_{\rm j}$ is the duration of the jet.
The minimum proton energy is $\varepsilon_{\rm p,m}=\Gamma_{\rm j}\varepsilon_{\rm p,m}'=\Gamma_{\rm j}(10m_{\rm p}c^2)$ and the maximum energy can be obtained by balancing the acceleration and cooling timescales,
\begin{equation}
\label{Eq:t_p_acc}
t_{\rm p,acc}'^{-1}>t_{\rm p,cool}'^{-1}=t_{\rm dyn}'^{-1}+t_{\rm syn}'^{-1}+t_{\rm p\gamma}'^{-1},
\end{equation}
where the acceleration timescale is $t_{\rm p,acc}'=\varepsilon_{\rm p}'/(ecB')$, the proton synchrotron cooling timescale is $t_{\rm p.syn}'=6\pi m_{\rm p}^4c^3/(m_{\rm e}^2\sigma_{\rm T}B'^2\varepsilon_{\rm p})$ and the photomeson cooling rate is \citep{Waxman-Bahcall(1997)},
\begin{equation}
\label{Eq:t_pgamma}
t_{\rm p\gamma}'^{-1}=\frac{c}{2\gamma_{\rm p}'^2}\int_{\bar{\varepsilon}_{\rm th}}^\infty d\bar{\varepsilon}_{\gamma}\sigma_{\rm p\gamma}\kappa_{\rm p\gamma}\bar{\varepsilon}_\gamma
\int_{\bar{\varepsilon}_\gamma/2\gamma_{\rm p}'}^\infty d\varepsilon_\gamma'\varepsilon_{\gamma}'^{-2}n'_\gamma(\varepsilon_\gamma'),
\end{equation}
where $\gamma_{\rm p}'=\varepsilon_{\rm p}'/m_{\rm p}c^2$ is the Lorentz factor of the proton, $\bar{\varepsilon}_\gamma$ is the photon energy in the proton rest frame, and $\bar{\varepsilon}_{\rm th}\approx 145 $ MeV is threshold energy for the photomeson production. $\sigma_{\rm p\gamma}$ and $\kappa_{\rm p\gamma}$ are the cross-section and inelasticity for photomeson production, respectively.
We use the fitting formulae based on GEANT4 in \citet{Kossov(2002)} for $\sigma_{\rm p\gamma}$.
For simplicity, we calculate the inelasticity using the following piecewise function:
$\kappa_{\rm p\gamma}=0.2$ for $\bar{\varepsilon}_\gamma<983$ MeV and $\kappa_{\rm p\gamma}=0.6$ for $\bar{\varepsilon}_\gamma>983$ MeV \citep{Atoyan-Dermer(2001)}\footnote{For more details on the theory and data for the inelasticity, see \citet{Stecker(1968)} and \citet{Patrignani-et-al.(2016)}.}.

\subsection{neutrino production}
Pions produced by the photomeson process decay into a muon and a muon neutrino. The spectrum of the muon neutrino can be expressed as,
\begin{equation}
\label{Eq: epsilon_nu_mu}
\varepsilon_{\rm \nu_\mu}^2n_{\rm \nu_\mu}\approx 1/8f_{\rm p\gamma}f_{\rm sup,\pi}\varepsilon_{\rm p}^2n_{\rm p},
\end{equation}
where $\varepsilon_{\rm \nu_\mu}\approx 0.05\varepsilon_{\rm p}$ is the muon neutrino energy, $f_{\rm p\gamma}=t_{\rm p,cool}'/t'_{\rm p\gamma}$ is the fraction of the proton energy converted into the pions \citep{Kimura-et-al.(2017)} and $f_{\rm sup,\pi}=1-\exp (-t_{\rm \pi ,cool}/t_{\rm \pi, dec})$ is the suppression factor due to pion cooling. $t_{\rm \pi,dec}=\gamma_\pi\tau_\pi$ is the decay timescale in the frame of pions, $\gamma_\pi=\varepsilon_\pi/m_\pi c^2$ is the Lorentz factor and $\tau_\pi=2.6\times 10^{-8}~$s. $t_{\rm \pi ,cool}^{-1}=t_{\rm \pi, syn}^{-1}+t_{\rm dyn}^{-1}$ is the cooling time and $t_{\rm \pi, syn}$ is the synchrotron cooling timescale for pions. Pions can decay into electron neutrinos and anti-muon neutrinos with the following energy spectra,
\begin{equation}
\label{Eq: epsilon_nu_e}
\varepsilon_{\rm \nu_{\rm e}}^2n_{\rm \nu_{\rm e}}\approx \varepsilon_{\rm \bar{\nu}_{\rm \mu}}^2 n_{\rm \bar{\nu}_{\rm \mu}}\approx 
1/8f_{\rm p\gamma}f_{\rm sup,\pi}f_{\rm sup,\mu}\varepsilon_{\rm p}^2n_{\rm p},
\end{equation}
where $\varepsilon_{\rm \nu_{\rm e}}\approx \varepsilon_{\rm \bar{\nu}_{\rm \mu}}\approx 0.05\varepsilon_{\rm p}$ are the energies of electron neutrinos and anti-muon neutrinos. $f_{\rm sup,\mu}=1-\exp (-t_{\rm \mu ,cool}/t_{\rm \mu, dec})$ is the suppression factor from muons, where $t_{\rm \mu, dec}=\gamma_\mu \tau_\mu$, $\gamma_\mu=\varepsilon_\mu/m_\mu c^2$ is the Lorentz factor of the muons and $\tau_\mu=2.2\times 10^{-6}~$s. $t_{\rm \mu ,cool}^{-1}=t_{\rm \mu, syn}^{-1}+t_{\rm dyn}^{-1}$ is the cooling time and $t_{\rm \mu, syn}$ is the synchrotron cooling timescale for muons.
Besides, neutrino oscillations cause a divergence between the spectrum observed on Earth and the source spectrum,
and the fluence is given by \citep{Harrison-et-al.(2002)},
\begin{equation}
\label{Eq: phi_nu_e}
\phi_{\rm \nu_{\rm e}+\bar{\nu}_{\rm e}}=\frac{10}{18}\phi_{\rm \nu_{\rm e}+\bar{\nu}_{\rm e}}^0
+\frac{4}{18}\left(\phi_{\rm \nu_{\rm \mu}+\bar{\nu}_{\rm \mu}}^0+\phi_{\rm \nu_{\rm \tau}+\bar{\nu}_{\rm \tau}}^0\right),
\end{equation}

\begin{equation}
\label{Eq: phi_nu_mu}
\phi_{\rm \nu_{\rm \mu}+\bar{\nu}_{\rm \mu}}=\frac{4}{18}\phi_{\rm \nu_{\rm e}+\bar{\nu}_{\rm e}}^0
+\frac{7}{18}\left(\phi_{\rm \nu_{\rm \mu}+\bar{\nu}_{\rm \mu}}^0+\phi_{\rm \nu_{\rm \tau}+\bar{\nu}_{\rm \tau}}^0\right),
\end{equation}
where $\phi_{\rm i}^0=n_{\rm i}/4\pi d_{\rm L}^2$ is the neutrino fluence at the source, $d_{\rm L}$ is the luminosity distance, and $\nu_{\rm \tau}$ ($\bar{\nu}_{\rm \tau}$) is the tau (anti-tau) neutrino.

\section{Results}\label{Sec:results}
In this section, we present the specific model parameters in Sec. \ref{sec: parameter} and then compute the radiation spectrum for seed photons (Sec. \ref{sec: seedPhoton}),  proton cooling timescale (Sec.  \ref{protons}), and neutrino emission (Sec. \ref{Sec: neutrino emissopn}).

\subsection{Parameter Setting} \label{sec: parameter}
We collect the specific parameter values for the SGRB jet in Table \ref{tab:parameter}, where $t_{\rm var}$ is the minimum variability timescale, which determines the radius of the internal dissipation region $r_{\rm diss}=2\Gamma_{\rm j}^2ct_{\rm var}=5\times 10^{13}$ cm. For the AGN disk and BCO parameters adopted in this work, we find that $r_{\rm diss}>r_{\rm ph,w}$, implying that AGN disk photons can still reach the jet interior when the cavity is filled with outflow material.

In addition, to facilitate a more intuitive comparison of the effects of the SMBH mass and the BCO location ($R_{\rm GRB}$), we perform calculations for three representative sets of parameters. Note that the necessary condition for efficient electron and proton acceleration in the internal dissipation region is the optical depth of the upstream shock $\tau_{\rm in}=n'\sigma_{\rm T}r_{\rm diss}/\Gamma_{\rm j}\lesssim 1$, where $n'=L_{\rm j,iso}/(4\pi r_{\rm diss}^2\Gamma_{\rm j}^2m_{\rm p}c^3)$ is the number density. According to Table \ref{tab:parameter}, $\tau_{\rm in}\approx0.08$, that is, the particles can be effectively accelerated.
To determine the photon number density of Eq. (\ref{eq: band}), we derive the peak energy $\varepsilon_{\gamma,p}=\Gamma_{\rm j}\varepsilon'_{\gamma,p}\approx 900$ keV using the relation in \citet{Qin-Chen(2013)}.

\subsection{Seed photons for p$\gamma$ processes} \label{sec: seedPhoton}
\begin{figure*}
\centering
\includegraphics [angle=0,scale=0.26] {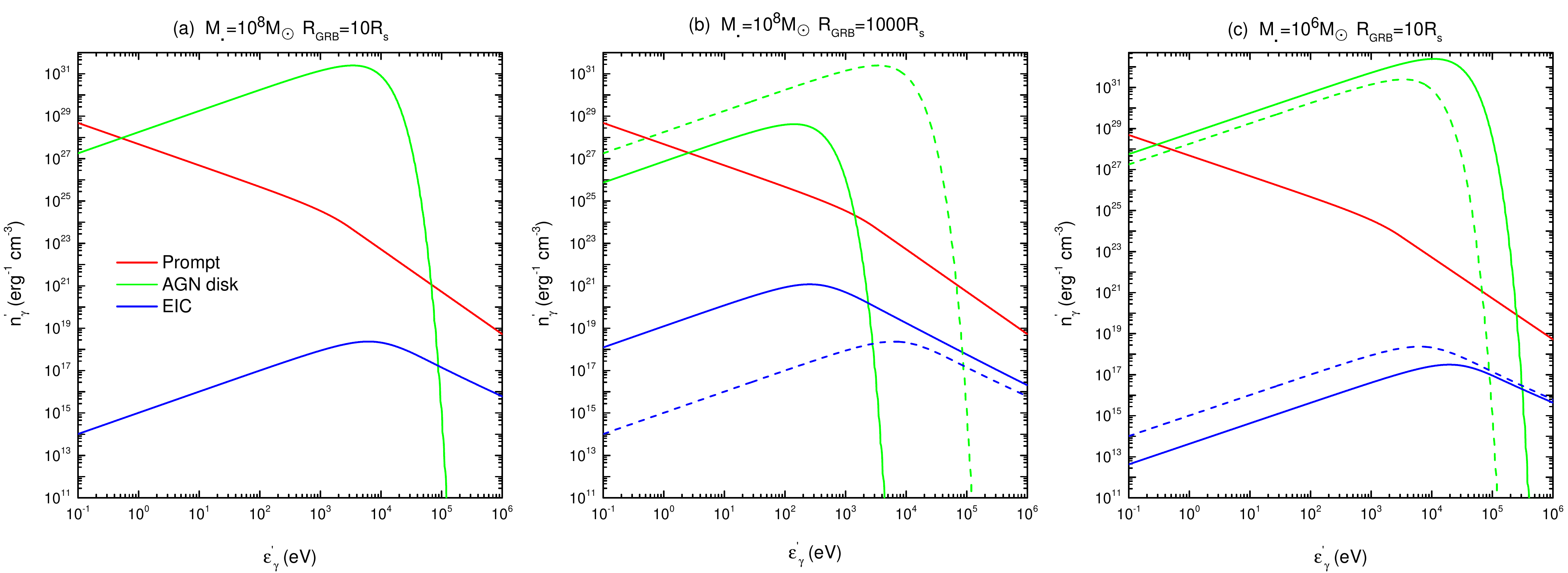}
\caption{The seed photon number density in the jet comoving frame for the p$\gamma$ reaction with $M_\bullet=10^8M_\odot$, $R_{\rm GRB}=10 R_{\rm s}$ (panel a), $M_\bullet=10^8M_\odot$, $R_{\rm GRB}=1000 R_{\rm s}$ (panel b), and $M_\bullet=10^6M_\odot$, $R_{\rm GRB}=10 R_{\rm s}$ (panel c). The red, green, and blue lines present the number densities of prompt, AGN disk, and EIC photons, respectively. In addition, to illustrate the effects of the SMBH mass and the BCO location, we overplot the results from panel (a) as dashed lines onto panels (b) and (c).
}
\label{fig: seedPhoton}
\end{figure*}

Fig. \ref{fig: seedPhoton} shows the seed photon number density distributions in the jet comoving frame, i.e., prompt (red line), EIC (blue line), and AGN disk (green line) photon fields.
In each panel, we label the specific $M_\bullet$ and $R_{\rm GRB}$ in the panel title. 
Furthermore, we verify that $\gamma\gamma$ annihilation is inefficient within the photon energy band under consideration. Therefore, $\gamma\gamma$ absorption does not alter the seed photon number density spectrum.

Since the prompt photon field is determined by the jet properties, the photon number density is the same in each panel.
In the jet comoving frame, the break energy is $\sim 3$ keV. The photon number density at the break is $\sim 5\times 10^{23}~\rm { erg^{-1}~cm^{-3}}$.
The photon number density increases as the photon energy decreases and reaches its maximum ($\sim 5\times 10^{28}~\rm { erg^{-1}~cm^{-3}}$) at the minimum photon energy ($\varepsilon'_{\gamma,m}=0.1$ eV) considered in our calculations.

The AGN disk photon number density spectrum varies with the adopted parameters and exceeds the prompt photon number density spectrum over certain energy ranges. In panel (a), the AGN disk photon number density spectrum exceeds the prompt photon number density spectrum between $0.5-6.5\times 10^4$ eV, and the peak number density is $\sim 2.5\times 10^{31}~\rm { erg^{-1}~cm^{-3}}$ at 4 keV.
With the BCO position $R_{\rm GRB}$ increasing, the photon number density spectrum of the AGN disk is lower, e.g., the spectrum exceeds the prompt photon number density spectrum between $2.5$ and $1.4\times 10^3$ eV, and the peak number density is $\sim 4\times 10^{28}~\rm { erg^{-1}~cm^{-3}}$ at 0.16 keV in panel (b).
Similarly, as the SMBH mass $M_\bullet$ decreases, the photon number density spectrum of the AGN disk is larger, e.g., the spectrum exceeds the prompt photon number density spectrum between $0.2-2.5\times 10^5$ eV, and the peak number density is $\sim 2.4\times 10^{32}~\rm { erg^{-1}~cm^{-3}}$ at 14 keV in panel (c). Therefore, variations in the BCO location have a more significant impact on the AGN disk photon number density spectrum than variations in the SMBH mass.

Under different parameter sets, the EIC photon number density spectrum exhibits a trend opposite to that of the AGN disk photon field, and the peak number density is $2\times 10^{18}~$, ($1\times 10^{21}$, $3\times 10^{17}$) $\rm { erg^{-1}~cm^{-3}}$ at 7 (0.28, 27) keV in panel a (b, c).
One should note that the EIC photon number density spectrum remains several orders of magnitude below the prompt photon number density spectrum across the entire energy range considered. Consequently, the contribution of EIC photons to neutrino production is neglected throughout subsequent calculations.

\subsection{Acceleration and Cooling of Protons}\label{protons}
\begin{figure}
\centering
\includegraphics [angle=0,scale=0.3] {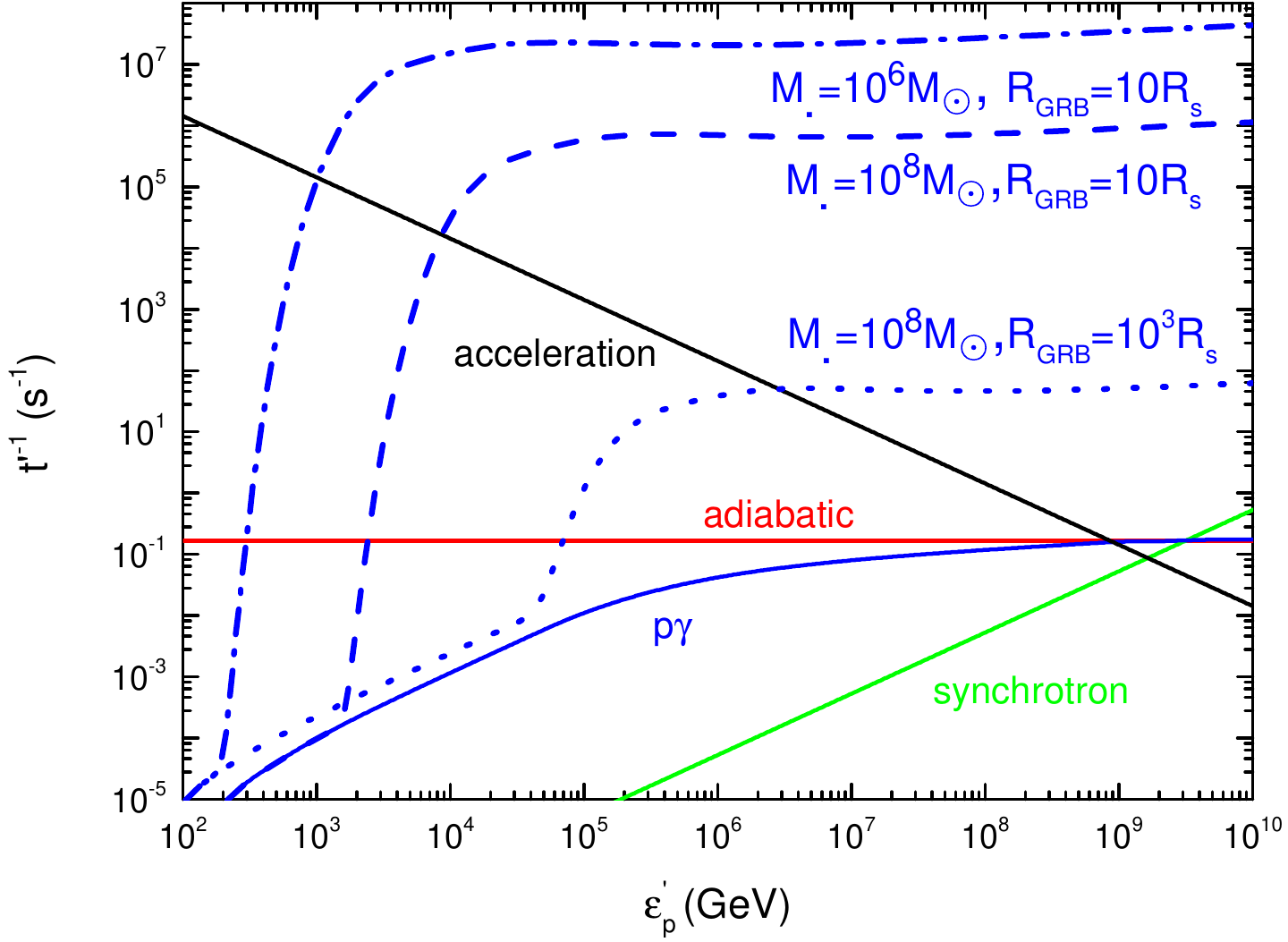}
\caption{The timescales for proton acceleration (black line), adiabatic cooling (red line), synchrotron radiation (green line), and the p$\gamma$ process (blue line). Besides, the timescales of the p$\gamma$ process for different SMBH masses and BCO positions are shown by different lines, e.g., $M_\bullet=10^8M_\odot$, $R_{\rm GRB}=10 R_{\rm s}$ (dashed line); $M_\bullet=10^8M_\odot$, $R_{\rm GRB}=10^3 R_{\rm s}$ (dotted line); and $M_\bullet=10^6M_\odot$, $R_{\rm GRB}=10 R_{\rm s}$ (dash-dotted line).
}
\label{fig: proton_timescale}
\end{figure}

Fig. \ref{fig: proton_timescale} presents the proton acceleration timescale (black line) and the cooling timescales for different cooling processes, such as adiabatic cooling (red line), synchrotron radiation (green line), and the p$\gamma$ process (blue line), where solid lines represent the case without AGN disk photons. In addition, we use different lines to present different parameters for $M_\bullet$ and $R_{\rm GRB}$, e.g., $M_\bullet=10^8M_\odot$, $R_{\rm GRB}=10 R_{\rm s}$ (dashed), $M_\bullet=10^8M_\odot$, $R_{\rm GRB}=10^3 R_{\rm s}$ (dotted), and $M_\bullet=10^6M_\odot$, $R_{\rm GRB}=10 R_{\rm s}$ (dash-dotted).

As shown in Fig. \ref{fig: proton_timescale}, when the AGN disk photon field is not included, adiabatic cooling dominates the proton energy losses for $\varepsilon_{\rm p}'\lesssim 10^9$ GeV, whereas synchrotron cooling dominates at higher energies. The p$\gamma$ cooling process, however, remains subdominant throughout the entire energy range considered. Once the AGN disk photon field is included, the p$\gamma$ cooling process becomes important and even dominates the proton energy losses for $\varepsilon_{\rm p}'\gtrsim 2~(75,~0.3)$ TeV for $M_\bullet=10^8M_\odot$, $R_{\rm GRB}=10 R_{\rm s}$ ($M_\bullet=10^8M_\odot$, $R_{\rm GRB}=10^3 R_{\rm s}$ and $M_\bullet=10^6M_\odot$, $R_{\rm GRB}=10 R_{\rm s}$). This is because the enhanced target photon density leads to a rapid increase in the photomeson interaction rate $f_{\rm p\gamma}$.

\subsection{Neutrino emission} \label{Sec: neutrino emissopn}
\begin{figure}
\centering
\includegraphics [angle=0,scale=0.3] {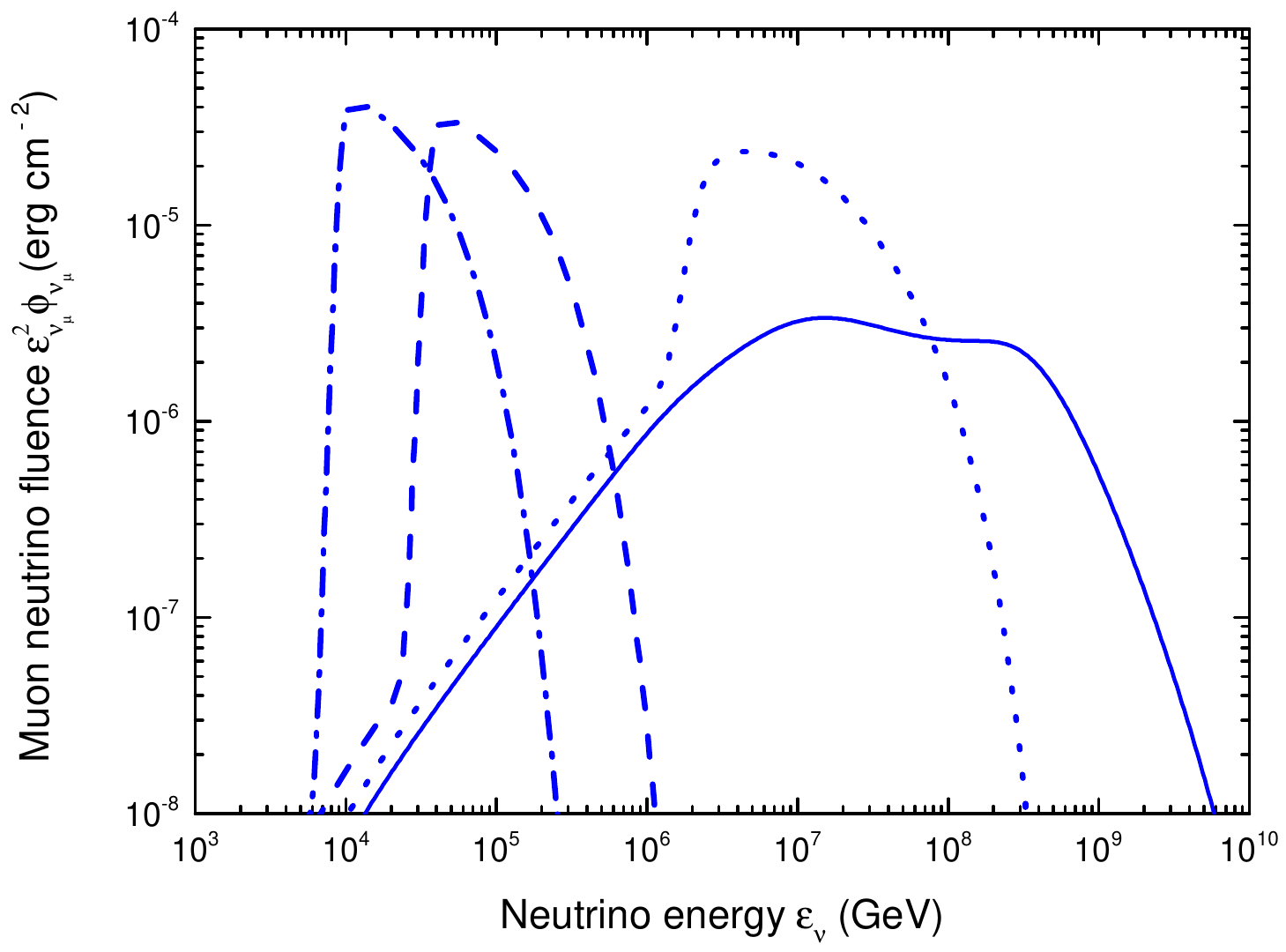}
\caption{The observable fluence of muon neutrino at $d_{\rm L}=300$ Mpc with different seed photon sources. The solid line is the case without AGN disk photons. Dashed, dotted, and dash-dotted lines represent the cases including the AGN disk photons for $M_\bullet=10^8M_\odot$, $R_{\rm GRB}=10 R_{\rm s}$; $M_\bullet=10^8M_\odot$, $R_{\rm GRB}=10^3 R_{\rm s}$; and $M_\bullet=10^6M_\odot$, $R_{\rm GRB}=10 R_{\rm s}$, respectively.}
\label{fig: neutrinoFlux}
\end{figure}

Fig. \ref{fig: neutrinoFlux} shows the observable fluence of muon neutrinos at $d_{\rm L}=300$ Mpc with different seed photons.
Similarly, we use the solid line to represent the case without the AGN disk photon field. Other lines show results including the AGN disk photons for different parameters, e.g., $M_\bullet=10^8M_\odot$, $R_{\rm GRB}=10 R_{\rm s}$ (dashed line), $M_\bullet=10^8M_\odot$, $R_{\rm GRB}=10^3 R_{\rm s}$ (dotted line), and $M_\bullet=10^6M_\odot$, $R_{\rm GRB}=10 R_{\rm s}$ (dash-dotted line).

As shown in Fig. \ref{fig: neutrinoFlux}, when the seed photon field consists only of prompt photons, the neutrino emission is concentrated in the energy range of $10^4-10^{10}$ GeV, with a peak fluence $3\times 10^{-6}~{\rm erg~cm^{-2}}$ at $1.5\times 10^7$ GeV.
The inclusion of the AGN disk photon field in the seed photon field leads to significant changes in the neutrino energy spectrum. Namely, the energy range of $10^4-10^{6},~10^4-3\times 10^{8},~\text{and}~5\times 10^3-2\times 10^5$ GeV, with a peak fluence $3.5\times 10^{-5}~2.4\times 10^{-5}, \text{and}~4.2\times 10^{-5}$ $~{\rm erg~cm^{-2}}$ at $4.6\times 10^4,~4.2\times 10^5,~\text{and}~1.2\times 10^4$ GeV for $M_\bullet=10^8M_\odot$, $R_{\rm GRB}=10 R_{\rm s}$ (dashed line), $M_\bullet=10^8M_\odot$, $R_{\rm GRB}=10^3 R_{\rm s}$ (dotted line), and $M_\bullet=10^6M_\odot$, $R_{\rm GRB}=10 R_{\rm s}$ (dash-dotted line), respectively.
Consequently, including the AGN disk photon field narrows the energy range of the emitted neutrinos while boosting their peak energy flux; this effect becomes significant for smaller $M_\bullet$ and $R_{\rm GRB}$.

\subsection{Neutrino Detection}\label{Sec: neutrino detection}
The detectable neutrino count ($N_\nu$) depend on both the neutrino fluence ($\phi_\nu$) and the detector effective area ($A_{\rm eff}$),
\begin{equation}
\label{Eq: N_nu}
N_\nu=\int \phi_\nu(\varepsilon_\nu)A_{\rm eff}(\varepsilon_\nu,\delta)d\varepsilon_\nu,
\end{equation}
where $\delta$ is the declination of the neutrino source. We use the effective area of IceCube for a 10-year point source search to calculate expected neutrino signal events \citep{IceCube-Collaboration-et-al.(2021)}. The effective area depends on neutrino energy and declination of the source, and it changes with the observational period due to instrumental variations. We adopt the effective area of IC86II in the calculation. 
Besides, the effective area of IceCube-Gen2 is $\sim 5$ times greater than that of IceCube \citep{Aartsen-et-al.(2021)}, so we scale IC86II to estimate that of IceCube-Gen2.
The probability that IceCube detects at least one neutrino can be expressed as \citep{Kimura-et-al.(2017), Matsui-et-al.(2023)},
\begin{equation}
\label{Eq: P_N_nu}
P (N_\nu \geqslant 1)=1-e^{-N_\nu},
\end{equation}

\begin{figure*}
\centering
\includegraphics [angle=0,scale=0.3] {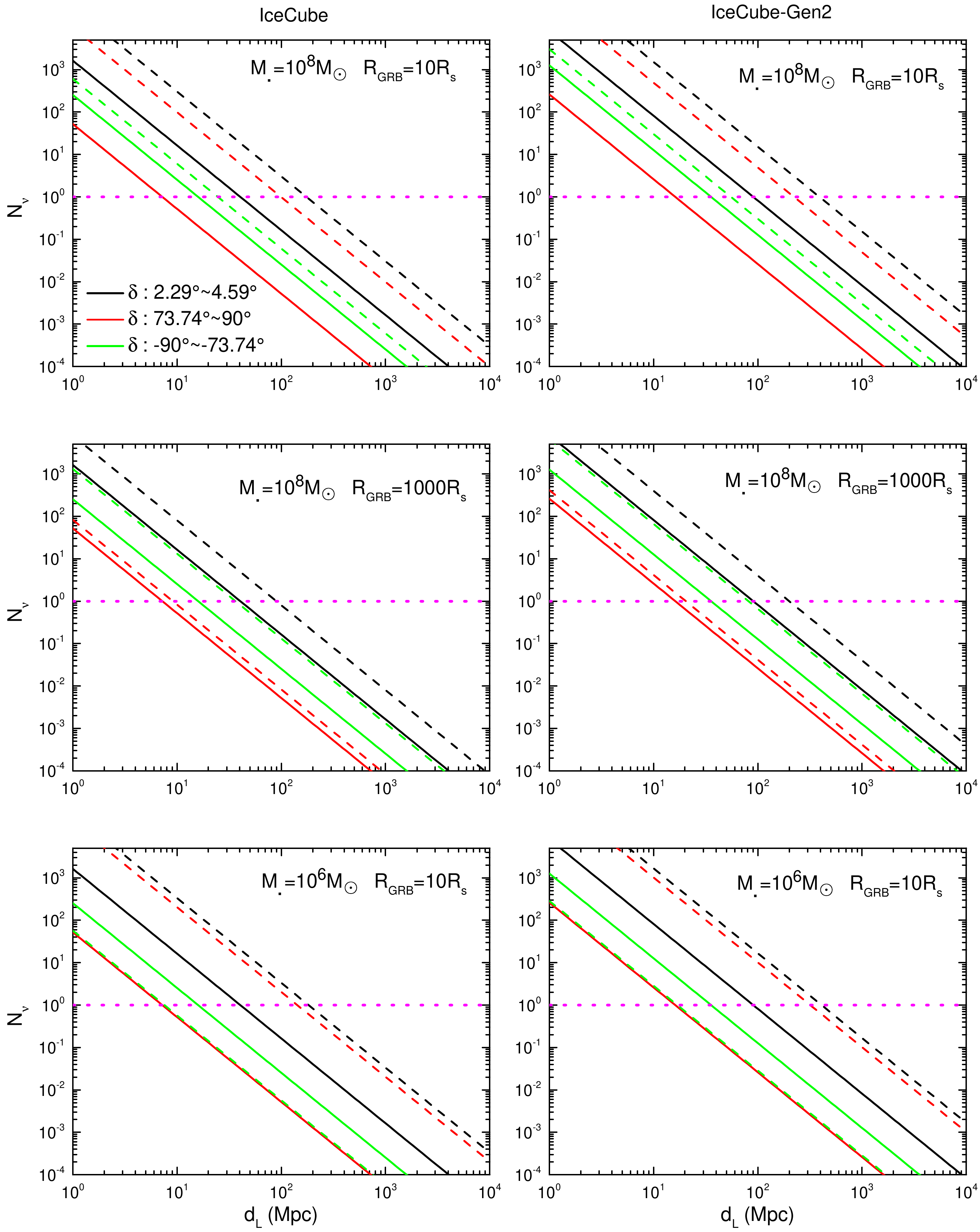}
\caption{The number of detected neutrinos varies with the luminosity distance $d_{\rm L}$. The left (right) column shows the result for the IceCube (IceCube-Gen2) detector. The top (middle, and bottom) row is the parameter for $M_\bullet=10^8M_\odot$, $R_{\rm GRB}=10 R_{\rm s}$ ( $M_\bullet=10^8M_\odot$, $R_{\rm GRB}=10^3 R_{\rm s}$; $M_\bullet=10^6M_\odot$, $R_{\rm GRB}=10 R_{\rm s}$).
The black (red, green) line is $\delta~\text{in}~ 2.29^\circ\sim 4.59^\circ$ ($73.74^\circ\sim90^\circ$, $-90^\circ\sim -73.74^\circ$). 
The solid and dashed lines represent the results obtained without and with the AGN disk photon field, respectively.
Besides, the magenta line represents $N_\nu=1$.
}
\label{fig: probability}
\end{figure*}

For subsequent calculations, we also choose three representative $\delta$ intervals to analyze the declination effect, i.e., $-90^\circ\sim -73.74^\circ$, $2.29^\circ\sim 4.59^\circ$ and $73.74^\circ\sim 90^\circ$. 
Specifically, near the south celestial pole, the effective area of the detector is dominant at $\varepsilon_\nu\gtrsim 10^6$ GeV; near the celestial equator, it remains large across a range of $10^2\lesssim\varepsilon_\nu\lesssim10^9$ GeV; whereas near the north celestial pole, it is dominant at $\varepsilon_\nu\lesssim 10^6$ GeV \citep[see Fig. 1 of][]{Ou-et-al.(2024)}. The detectable neutrino counts of IceCube (Gen2) are shown in the left (right) column of Fig. \ref{fig: probability}.
The top, middle, and bottom panels are the results for $M_\bullet=10^8M_\odot$, $R_{\rm GRB}=10 R_{\rm s}$,  $M_\bullet=10^8M_\odot$, $R_{\rm GRB}=10^3 R_{\rm s}$, and $M_\bullet=10^6M_\odot$, $R_{\rm GRB}=10 R_{\rm s}$, respectively.
The black, red, and green lines represent $\delta: 2.29^\circ\sim 4.59^\circ$, $73.74^\circ\sim90^\circ$, and $-90^\circ\sim -73.74^\circ$, respectively. The solid and dashed lines represent the results obtained without and with the AGN disk photon field, respectively.
The magenta line represents neutrino counts $N_\nu=1$. Hereafter, we denote the maximum luminosity distance for detecting at least one neutrino as $d_{\rm N_{\rm \nu}=1}$.

As shown in Fig. \ref{fig: probability}, 
for the case without an AGN disk photon field, the detection distance depends solely on $\delta$.
For $\delta~\text{in}~2.29^\circ\sim 4.59^\circ$ ($73.74^\circ\sim90^\circ$, $-90^\circ\sim -73.74^\circ$), we have $d_{\rm N_{\rm \nu}=1}=$ 40 (7, 15) Mpc for IceCube, and $d_{\rm N_{\rm \nu}=1}=$90 (20, 35) Mpc for IceCube-Gen2.
As revealed by these results, sources located near the celestial equator exhibit the maximum detectable distance, which is due to a combination of a broad emission bandwidth and a favorable detector effective area.

When the AGN disk photon field is included, $d_{\rm N_{\rm \nu}=1}$ depends not only on $\delta$, but also on $M_\bullet$ and $R_{\rm GRB}$:  
(1) For $M_\bullet=10^8M_\odot$, $R_{\rm GRB}=10R_{\rm s}$, and a source located at $\delta: 2.29^\circ\sim 4.59^\circ$ ($73.74^\circ\sim90^\circ$, $-90^\circ\sim -73.74^\circ$), we have $d_{\rm N_{\rm \nu}=1}=$170 (100, 25) Mpc for the IceCube, and $d_{\rm N_{\rm \nu}=1}=$390 (220, 55) Mpc for the IceCube-Gen2.
(2) For $M_\bullet=10^8M_\odot$ and $R_{\rm GRB}=1000R_{\rm s}$, and a source located at $\delta~\text{in}~2.29^\circ\sim 4.59^\circ$ ($73.74^\circ\sim90^\circ$, $-90^\circ\sim -73.74^\circ$), we have $d_{\rm N_{\rm \nu}=1}=$90 (9, 35) Mpc for the IceCube, and $d_{\rm N_{\rm \nu}=1}=$200 (20, 85) Mpc for the IceCube-Gen2.
(3) For $M_\bullet=10^6M_\odot$ and $R_{\rm GRB}=10R_{\rm s}$, and a source located at $\delta~\text{in}~ 2.29^\circ\sim 4.59^\circ$ ($73.74^\circ\sim90^\circ$, $-90^\circ\sim -73.74^\circ$), we have $d_{\rm N_{\rm \nu}=1}=$180 (140, 7) Mpc for IceCube, and $d_{\rm N_{\rm \nu}=1}=$400 (300, 15) Mpc for IceCube-Gen2.
After including the AGN disk photon field, the peak energy of the neutrino spectrum shifts toward lower energies, thereby significantly extending $d_{\rm N_{\rm \nu}=1}$ near the celestial equator and the north celestial pole, particularly around the celestial equator.


\begin{table}
\centering \caption{GRB jet parameters}
\begin{tabular}{lc}
\hline
parameter                          & value     \\ \hline
$\Gamma_{\rm j}$                   & 300       \\
$L_{\rm j,iso}~({\rm erg~s^{-1}})$ & $10^{53}$ \\
$t_{\rm j}~(s)$                    & 1         \\
$\epsilon_{\rm e}$                 & 0.1       \\
$\epsilon_{\rm B}$                 & 0.01      \\
$\epsilon_{\rm p}$                 & 0.3       \\
$t_{\rm var}~(s)$                  & 0.01      \\ \hline
\end{tabular}
\label{tab:parameter} 
\end{table}

\section{Conclusions and Discussions}\label{sec: conclusion}
In this paper, we investigate the neutrino emission characteristics of SGRB jets propagating within AGN disk cavities, and evaluate the detection capabilities of IceCube for such neutrino sources.
Taking into account three SMBH masses, we find that the radius of the cavity does not exceed the disk photosphere radius for $R_{\rm GRB}\gtrsim 10^3 R_{\rm s}$ in an AGN disk with $M_\bullet=10^6M_\odot$.
Three seed photon fields are considered as the targets for neutrino production through photomeson interactions of accelerated protons in this work, i.e., prompt photons, EIC photons, and AGN disk photons. Since the number density of EIC photons is significantly lower than that of prompt photons, the contribution of EIC photons is omitted in our calculations of the neutrino spectra. 

Our results indicate that due to the high number density, the AGN disk photons play a crucial role in shaping the neutrino energy spectrum. Specifically, the peak emission is shifted toward lower energy bands. Such effects become more prominent as $M_\bullet$ or $R_{\rm GRB}$ decreases. 
Due to the contribution of AGN disk photons, IceCube's maximum detection distance for at least one SGRB neutrino is extended by a factor of several to tens for sources near the celestial equator and the north celestial pole. Conversely, for sources near the south celestial pole, this enhancement is negligible, and the distance can even be reduced.
Furthermore, we note that if the jet Lorentz factor is low and the SMBH mass is large (e.g., $\Gamma_{\rm j}\lesssim 100$ and $M_\bullet\gtrsim 10^8M_\odot$), the dissipation radius may be smaller than the outflow photospheric radius. Therefore, the current model is not directly applicable, and a detailed study of neutrino emission in this regime will be deferred to future work.

We anticipate that the new-generation neutrino detectors will possess higher sensitivity and larger effective areas to search for the neutrino radiation generated by SGRB jets in the AGN disk and to validate our model, e.g., Baikal-GVD \citep{Avrorin-et-al.(2011)}, IceCube-Gen2 and KM3NeT \citep{Adrian-Martinez-et-al.(2016)}.
Besides, searching for neutrinos from sGRB jets in AGN disks can help analyze jet composition and the central engine \citep{Ou-et-al.(2024),Lei2026}.
We adopt the effective temperature as the seed photon temperature in the calculation. However, the temperature in the deeper regions of the AGN disk is higher by about one order of magnitude (see Fig.\ref{fig:AGNdiskprofile_1}). 
If internal dissipation takes place deep inside the AGN disk, the temperature of the seed photons is significantly higher. The result in Fig.\ref{fig: neutrinoFlux} indicates that the neutrino emission softens to lower energy bands without a notable change in the peak spectral flux in this case, thereby yielding a slight enhancement of $d_{\rm N_{\rm \nu}=1}$ around the celestial equator and the north celestial pole. 
In addition to AGN disk photons, 
the wind from the accretion disk around the BCOs, together with the photons from the BCOs disk, could also provide additional seed photons.
The energy density of the BCO-disk photons at the dissipation radius can be estimated as $u_{\rm w}\approx L_{\rm w}/(4\pi r_{\rm diss}^2c)=1.95\frac{M\bullet}{10^6M_\odot}~{\rm erg~cm^{-3}}$, and the energy density of the AGN disk photons is $u_{\rm AGN}=aT^4$, where $a$ is the radiation constant.
We find that the photon energy density of the outflow is lower than that of the AGN disk photons, except for $R_{\rm GRB}\gtrsim 1000R_{\rm s}$ and $M_\bullet\gtrsim 10^7 M_\odot$.
Although the current analysis does not include the effect of BCO-disk photons, our results indicate that when these photons become important, the peak energy of the neutrino spectrum would shift to lower energies and further increase $d_{\rm N_{\rm \nu}=1}$ near the directions of the celestial equator and the north celestial pole.
We plan to investigate these issues in detail in future work.


The SGRBs in the AGN disk are ideal sources for multi-messenger observations. 
In addition to neutrino and EM emission from the jet in the research, before the BCO merger, the wind generated by accretion interacts with the material in the AGN disk to produce EM and neutrino radiation \citep{Wang-et-al.(2021a), Kimura-et-al.(2021), Zhou-et-al.(2023)}.
Furthermore, the evolution of a single compact star during the AGN disk can also generate a wind and accompanying radiation~\citep {Chen-et-al.(2023)}. GW observations will be crucial for determining the EM and neutrino radiation from the pre- or post-merger systems. Therefore, we expect that the next generation of space-based GW detectors can capture the GW signal within the BCO inspiral, e.g., Laser Interferometer Space Antenna \citep[LISA,][]{Amaro-Seoane-et-al.(2017)} and TianQin \citep{Luo-et-al.(2016)}.


\begin{acknowledgements}
We are very grateful to Chichuan Jin for helpful discussions, and the anonymous reviewer for his constructive comments and suggestions. 
This work is supported by the National Key R\&D Program of China (No. 2023YFC2205901), and the National Natural Science Foundation of China under grants 12473012, 12533005, 12003007 and the Fundamental Research Funds for the Central Universities (No. 2020kfyXJJS039). W.H.Lei. acknowledges support from the science research grants from the China Manned Space Project with NO.CMS-CSST-2021-B11.
\end{acknowledgements}

\bibliography{ref}{}

\end{document}